\documentclass[aps, prl, reprint]{revtex4-2}
\usepackage{graphicx}
\usepackage{color}
\usepackage{verbatim}
\usepackage{natbib}   
\usepackage{makecell}
\usepackage{multirow}
\usepackage{tabularx}
\usepackage[svgnames]{xcolor}
 \usepackage{hyperref}
 \usepackage{amsmath}
 \usepackage{amssymb}
 \usepackage{booktabs}
 \usepackage[normalem]{ulem}

 \usepackage{verbatim}

\newcommand{\lamav}{\textless$\lambda$\textgreater ~}

\newcommand{\archive}[1]{\href{https://nifit.llnl.gov/viewer/browseShot.action?shotId=#1-999}{Archive Viewer}} 

\hypersetup{
  colorlinks,
  citecolor=Violet,
  linkcolor=Black,
  urlcolor=Blue,
  linkcolor=Blue}

\begin{document}
\title{Determining the Self-Similar Stage of the Rayleigh-Taylor Instability via LLNL's NIF Discovery Science Experiments}
\author{A. Shimony$^{1, 2}$, C.M. Huntington$^{3}$, K.A. Flippo$^{4}$, Y. Elbaz$^{5}$, S.A. MacLaren$^{3}$, D. Shvarts$^{1, 6}$ and G. Malamud$^{1, 2}$}

\affiliation{$^{1}$Nuclear Research Center Negev, Beer Sheva 84190, Israel \\
$^{2}$University of Michigan, Ann Arbor, MI 48109, USA \\
$^{3}$Lawrence Livermore National Laboratory, Livermore, CA 94550, USA \\
$^{4}$ Los Alamos National Laboratory, Los Alamos, NM 87545, USA \\
$^{5}$Israel Atomic Energy Commission, Tel Aviv 61070, Israel \\
$^{6}$Ben Gurion University of the Negev, Beer Sheva 84105, Israel
}

\begin{abstract}

We report a novel measurement of the late time self-similar growth constant $\alpha_B$ of the Rayleigh-Taylor instability (RTI) from controlled initial perturbations. To this end, we have developed a laser-driven experiment, fielded on the National Ignition Facility (NIF), to explore deeply non-linear, multimode hydrodynamic growth of a planar interface. The measured value is $\alpha_B=0.038\pm 0.008$, which is consistent with previously reported 2D simulations and closer to previously reported 3D simulations. This resolved the known discrepancy between experiments and simulations of RTI.

\end{abstract}
\maketitle

The Rayleigh-Taylor instability (RTI) evolves when a light fluid pushes a heavy fluid  \cite{rayleigh1882investigation,taylor1950instability} and ubiquitous in physical systems of substantial different length scales like Inertial Confinement Fusion (ICF) capsules \cite{emery1982rayleigh} and supernovae \cite{guzman2009non}. In the case of a narrowband multimode initial perturbation, the RTI reaches the regime of self-similar evolution after $G\approx 3$ \cite{elbaz2018modal}, where $G$ is the number of bubble merger generations. At this stage, the magnitudes of the spike and bubble ($h_S$ and $h_B$, respectively) are given by $h_{S/B} = \alpha_{S/B}Agt^2$, where $g$ is a constant acceleration at the unstable interface, $A=\frac{\rho_2-\rho_1}{\rho_1+\rho_2}$ is the Atwood number ( $\rho_1$ and $\rho_2$ are the densities of the lighter fluid and the density of the heavier fluid, respectively), $t$ is the evolution time, and $\alpha$ is a constant that describes the growth, which may be different for the spike and bubble. The value of $\alpha_B$ governs the mixing process and has been a topic of research for decades. 

Despite active research, significant discrepancies remain between theoretical, computational, and experimental calculations of $\alpha_B$. Recent theoretical model results suggest a value of $\sim 0.05$ for 3D immiscible fluids \cite{elbaz2018modal}. This result is consistent with LEM experiments \cite{dimonte2000density} and bubble-merger models \cite{alon1995power, oron2001dimensionality}, but a factor of $\sim$2 higher than the results of full numerical 3D simulations \cite{dimonte2004comparative,zhou2017rayleigh}. This last result is claimed to be attributed to the entrainment of the two fluids in the bubble caused by a 3D small scale turbulent flow, which reduces the effective Atwood number by $\sim$2 for miscible fluids, as shown in full numerical simulations \cite{shimony2018density}.

To resolve the discrepancy, introduced above, between experiments and simulations, a precise comparison between them should be made. The initial perturbations in most of the past simulations were of a narrowband, so that long wavelengths were absent in the initial growth and the resulted evolution of the mixing zone was of bubble merger dynamics. However, in past experiments, the control over the initial perturbations was limited and initial long wavelengths could affect growth and increase the measured value of $\alpha_B$, as demonstrated by simulations \cite{youngs2013density}. In this work, we present a novel experiment, with a controlled narrowband initial perturbation \cite{malamud2014conceptual}, which was performed on the National Ignition Facility (NIF). The experimental results were compared to a model and to a hydrodynamic simulation.

Several aspects of the present experimental configuration were adapted from similar hydrodynamics studies on the National Ignition Facility (NIF), including the laser drive, structure of the ablator \cite{kuranz2018high, nagel2017platform}, low-density foam configuration \cite{nagel2017platform, huntington2018ablative}, and x-ray source \cite{flippo2016late, nagel2017platform}. Essential elements are described below; further details can be found in the relevant publications.

\begin{figure}
\includegraphics[width = 3.375in]{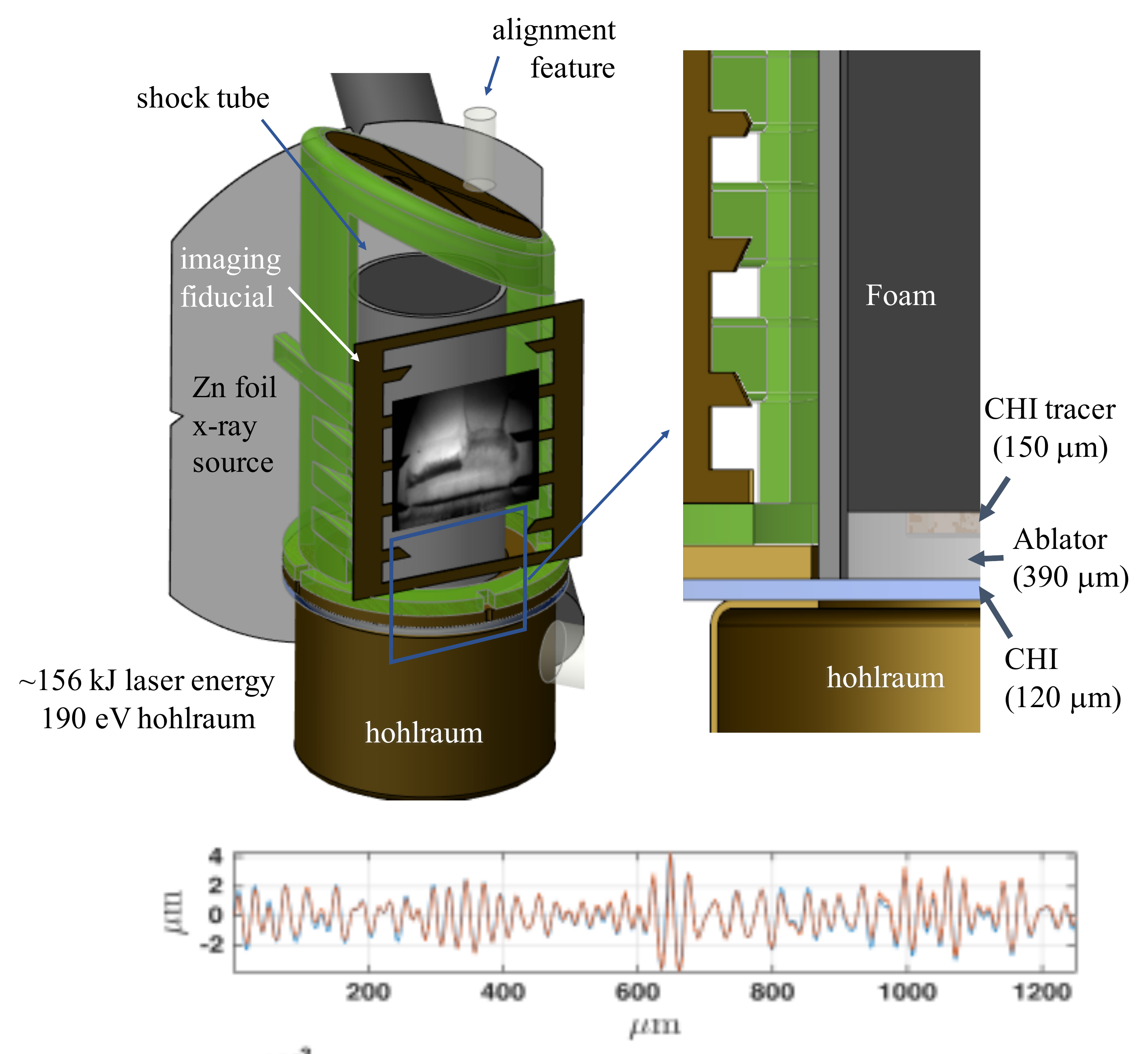}
\caption{The main components of the experiment---the CHI / PAI ablator with the CHI tracer and the low-density foam---are contained within a Be tube and mounted above the Au hohlraum. A printed plastic component (shown in green in this figure) is used to accurately position the Zn x-ray source foil and the Au fiducial on opposite sides of the physics package. The pre-imposed ripple pattern that is machined across the face of the plastic at the plastic-foam interface is shown below the target.}
\label{fig:target}
\end{figure}

The unstable interface under investigation consisted of a plastic disk at a density of 1.43 $g/cm^3$, mated to a foam cylinder. As shown in Fig. \ref{fig:target}, the portion of the plastic facing the hohlraum and served as the ``ablator'' was plastic doped with a fraction of iodine (C$_{50}$H$_{47}$I$_3$, ``CHI''), which served to stop moderate-energy x-rays (Au M-band $>$ 2 keV) from reaching the interface. Following the CHI ablator, a layer of polyamide-imide plastic (PAI, C$_{22}$H$_{14}$N$_2$O$_3$) formed the bulk of the dense interface, except for a 300 $\mu$m wide strip of the CHI that spanned half of the target width at the interface with the foam. While the PAI was largely transparent to the diagnostic x-rays, the purpose of this tracer strip was to absorb the x-rays and provide contrast in the image, highlighting the region location of the dense layer as it mixed into the low-density foam.

The same x-ray opaque tracer-layer concept that was used in the plastic was employed in the foam. Most of the foam cylinder was machined from carbon resorcinol foam (CRF) at an initial density of approximately 0.08 $g/cm^3$. In a groove machined half-way through the cylinder, a 300 $\mu$m wide strip of carbon foam with a nickel dopant (C$_{88}$Ni$_{12}$) \cite{huntington2020split}, also nominally 0.08 $g/cm^3$, was inserted. The nickel K-edge at 8.3 keV strongly absorbed the 9 keV diagnostic x-ray signal, providing a contrasting dark layer where the doped foam was mated to the PAI plastic.  

Indirect-drive, where laser energy is used to generate a uniform x-ray bath in a gold hohlraum, was used to produce the acceleration at the unstable interface.  The laser heating the hohlraum with approximately 158 kJ of laser energy in a $\sim6$ ns effective pulse. The hohlraum reached a peak temperature of approximately 230 eV, which served to ablate material from the first layer of iodinated plastic layer and drive a shock into the ablator. As the laser shuts down, the shock is no longer supported, and decays into a blast wave.

As the blast wave crosses the plastic-foam interface, it imposes an impulsive acceleration, resulting initial small growth via Richmyer-Meshkov instability (RMI), immediately followed by deceleration, driving the interface to the RTI unstable regime for a period of several 10's of ns, until the interface slows \cite{malamud2014conceptual}. As a result of the blast wave density and pressure profiles, an undesired decompression effect is introduced, stretching the mixing zone. This effect should be treated in the analysis along with the RMI effect, in order to achieve a clear understanding of the instability. We address this further below.

\begin{figure}
\includegraphics[width = 3.375in]{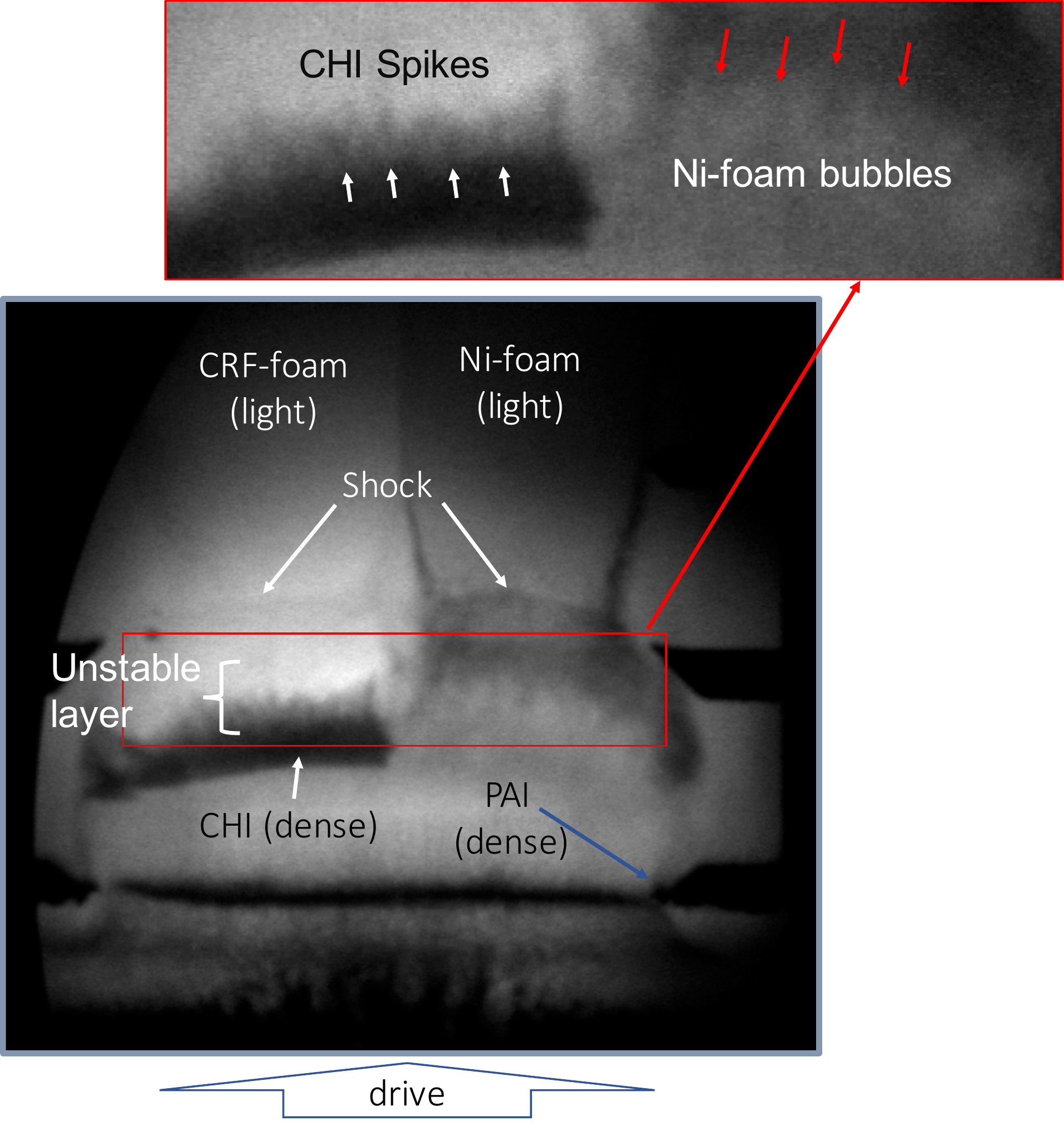}
\caption{An image from a target with 2D initial conditions (\lamav = 20 - 40 $\mu$m), taken 45 ns after the start of the laser drive. The inset image highlights the region where the spikes of dense material (CHI on the left and PAI on the right) are penetrating the foams. The shock is ahead (above) this unstable interface; its position is measured accurately using the fiducial bars on the edges of the image. The position the tracer is measured using the sharp interface with the CHI layer.}
\label{fig:image}
\end{figure}

The initial conditions at the plastic-foam interface sets the initial growth rate of the instability, and as such has an important role in the observed, late-stage mixing zone width. This campaign employed four distinct types of initial conditions, predicted to arrive to G$\approx$3 for the specific design at hand: 
\begin{enumerate}
\item 2D with average wavelength \lamav in a band 10 - 20 $\mu$m, 
\item 2D with \lamav = 20 - 40 $\mu$m, 
\item 3D with \lamav = 20 - 40 $\mu$m, where the 3D pattern was created by cutting the same 2D pattern in orthogonal directions, producing and ``egg crate'' pattern, and 
\item an additional 2D pattern with \lamav = 20 - 40 $\mu$m, but with a different ablator thickness than type (1).
\end{enumerate}

We note, that all these initial patterns were implemented with a flat spectrum, i.e. no dependence of the initial amplitude on the wave number (see also in \cite{malamud2014conceptual}).

\begin{figure}
\includegraphics[width = 3.375in]{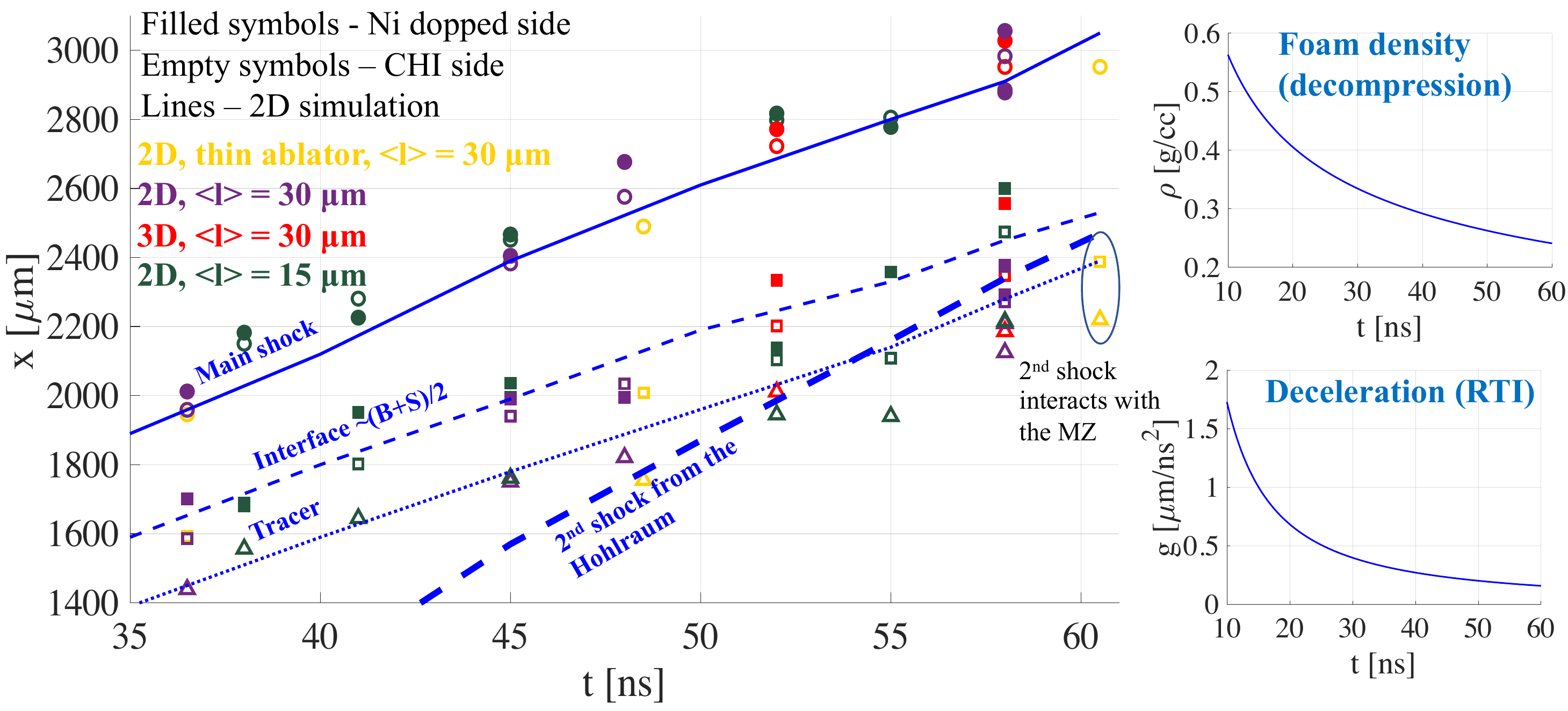}
\caption{1D position trajectories of the shock, interface, bubbles and the spikes fronts, tracer and the second shock from the hohlraum \cite{nagel2017platform}. The experimental positions of the main shock, the main interface and the tracer are given by circle, square and triangle symbols, respectively. The density and the acceleration are also presented as a function of time in the inserts.}
\label{fig:trajectories}
\end{figure}

The main diagnostic was X-ray radiography, producing images of the main physical features at different times. For each image, calibrated points on the target were recorded in the image and permitted absolute and relative measurements of features in the shock tube. An example image is shown in Fig. \ref{fig:image}, with relevant features labeled. From each image the location of the shock front, spike, bubble, and tracer were measured. Since only one image was recorded from each target, the time sequence was collected for each of the initial conditions described above. We note, that shot to shot variance in laser energy ($\sim$8\%) and target dimensions ($\sim$7\%) and density ($\sim$5\%) was taken into account using detailed 1D and 2D LASNEX simulations \cite{zimmerman1978lasnex}, allowing a calibration of the experimental results, as presented in Fig. \ref{fig:trajectories}. This figure shows good agreement between the simulation and the experimental results. It can be seen by the positions of the main shock, the main interface, and the tracer. This result justifies the usage of $g(t)$, obtained from the LASNEX simulations, in model and the simulations for the evolution of the hydrodynamic instabilities. Two notable exceptions are the latest shot, which was probably affected by the second shock wave from the hohlraum \cite{nagel2017platform}, as suggested from the 2D LASNEX simulation, and the 55ns shot, which seems by most measures (including the mix width, presented below) to be unexplainably weaker than the other shots. Both shots are presented but excluded from the analysis below.
 
For a simpler analysis of the evolution of the hydrodynamic instabilities and a further treatment of the shot-to-shot variation, we calculated the Read integral,

\begin{align}
x_{Read} = A\left(\int{\sqrt{g(t)}dt}\right)^2,
\end{align}
which provides a length scale for the distance traveled for each experimental configuration \cite{read1984experimental}. Without the effects of decompression and RMI, the asymptotic slope of the mixing zone as a function of $x_{Read}$ would be equal to $\alpha_S+\alpha_B$.

\begin{figure}
\includegraphics[width = 3.375in]{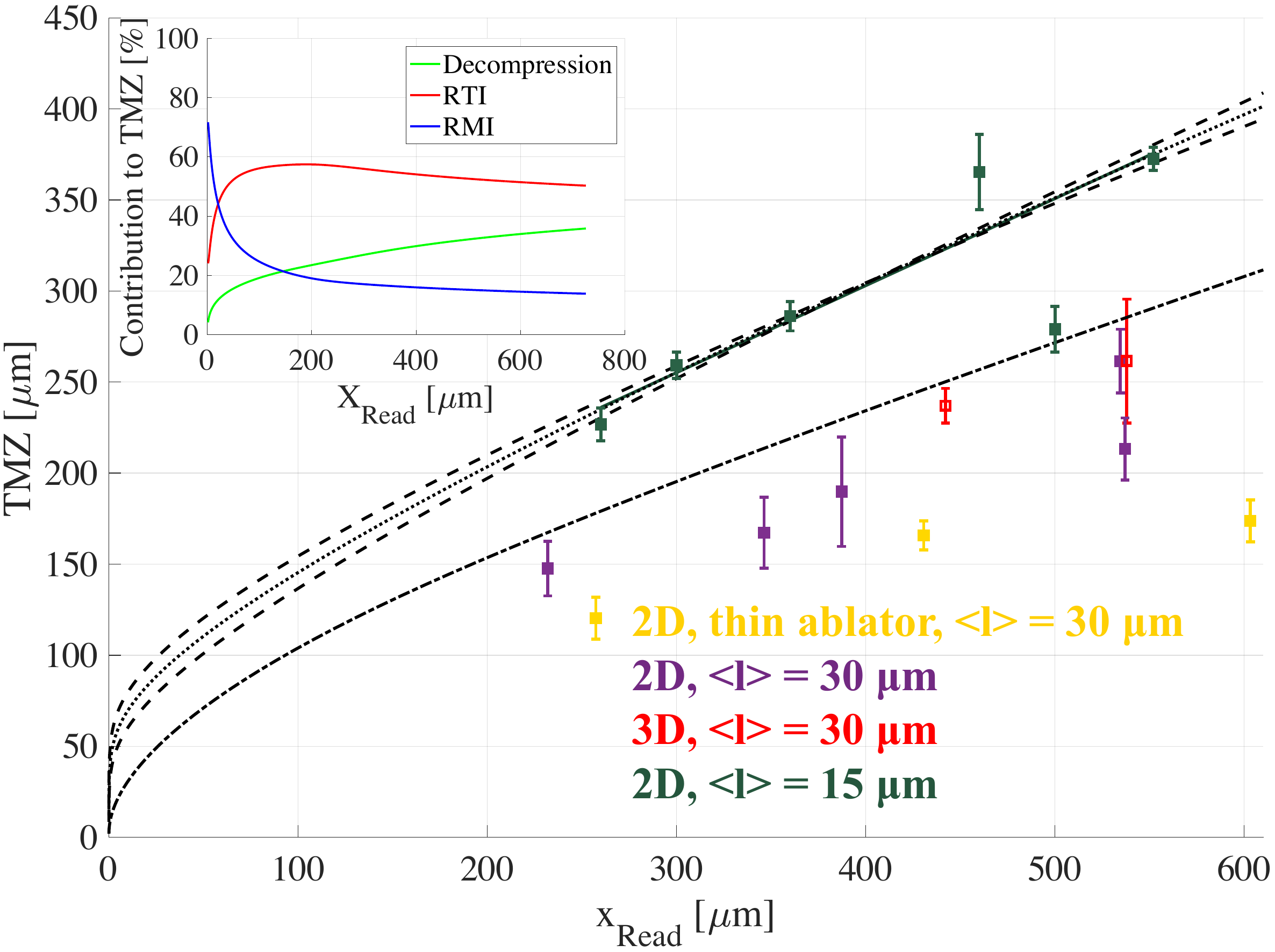}
\caption{The experimental data with errorbar for each group of initial perturbation together with model (lines) predictions for different cases. Dash-dotted line: $\alpha_B=0.03$ and $\left<\lambda\right>=30\mu m$, dashed line: $\alpha_B=0.03;0.046$ and $\left<\lambda\right>=15\mu m$, dotted line:  $\alpha_b=0.038$ and $\left<\lambda\right>=15\mu m$. Insert: the contributions of the main effects (RT instability, RM instability and decompression), predicted by the model.}
\label{fig:data}
\end{figure}

In Fig \ref{fig:data}, the mix width extracted from each shot is presented  as a function of  $X_{Read}$, grouped by initial condition type as enumerated previously (i.e. 2D$\backslash$3D). Several trends are immediately apparent from the data. First, the smallest initial wavelengths (2D, $\left<\lambda\right>=15\mu$m) exhibit the largest initial growth rate, and have reached the largest mix widths at the times probed. This is consistent with an initial growth rate proportional to the the wavenumber $k$ for RMI, considering that the initial amplitudes are very close for different initial wavelengths (i.e. due to manufacturing limitations). Additionally, as expected, 3D patterns achieve larger amplitudes than equivalent 2D patterns (i.e. $\left<\lambda\right>=30\mu$m). We attribute this to the inherent difference in wave number for identical wavelengths in 2D and 3D RT. 

An important conclusion from the data is, that growth rate of each of individual datasets is similar. Recall, this growth rate is a combination of several factors. Initially, the interface grows via a combination of RMI and RTI instability. The RTI growth dominates the system at intermediate times, as the RMI mechanism saturates and before target decompression becomes significant. At late times, target decompression stretches the mixing zone, an effect which can be mistaken for instability growth and must be deconvolve for the analysis.

In order to better understand the contribution of each physical mechanism to the total mix width a simple buoyancy-drag (BD) model can be implemented. Such a model accounts for non-linear RTI via a force-balance relationship (first and second terms in eq. \ref{BD1}), and decompression via a divergence term (last expression on the right side of eq. \ref{BD1}) \cite{srebro2003general,oron2001dimensionality, miles2004bubble, dimonte2000spanwise, huntington2018ablative}, and Eq. \ref{BD2} supplies a physical lower limit to $\left<\lambda\right>$ and a self-similar form for the late time solution. The initial shock contribution to RMI  is introduced to the model as an initial condition on the growth rate. The equations of the BD model, used in this work are,
\begin{gather}
\label{BD1}
\left(\rho_{1,2}+C_a\rho_{2,1}\right)\frac{du_{B,S}}{dt}= \\
\nonumber\left(\rho_2-\rho_1\right)g\left(t\right)-C_d\rho_{2,1}\frac{u_{B,S}^2}{\left<\lambda\right>}+\frac{d}{dt}\left(\frac{h_{B,S}}{\rho_{1,2}}\frac{d\rho_{1,2}}{dt}\right) \\
\label{BD2}
\left<\lambda\right>=\max\left(\left<\lambda_0\right>,\frac{h_B}{b}\right)
\end{gather}

\noindent where $u_{b,s}$ is the bubble's/spike's velocity, $C_d$ is the drag coefficient, $C_a$ is the added mass coefficient and $b$ is the ratio between the amplitude of the bubbles and their average wavelength, which is a free parameter of the model and is a one-to-one function of $\alpha_B$. We assumed that at the self-similar stage, the initial conditions are forgotten, so that $C_d=2\pi,6\pi$ and $C_a=1,2$, which are the values for the 3D and 2D cases, respectively \cite{srebro2003general}. Integrating over eq. \ref{BD1} yields the bubble and spike front velocities, and a second integral yields their amplitudes.

In addition to the experimental data discussed above, Fig. \ref{fig:data} presents the model predictions for $\left<\lambda\right>_0=30\mu m$, initial amplitude of 1$\mu m$ and $\alpha_B=0.03$. The relative contributions of each of the terms of the model is presented in the insert of Fig. \ref{fig:data} for this case. At $X_{Read}$ = 600 $\mu m$, slightly beyond the largest growth measured in the experiments, more than half of the total growth predicted by the BD model is a product of RTI. The addition of a RMI initial velocity was calculated from the Meyer-Blewett formula \cite{meyer1972numerical} $u_{RM}=\frac{2\pi}{\left<\lambda\right>}\Delta u \left(a_0+a_{ps}\right)/2
$, where $u_{RM}$ is the initial velocity of the bubbles due to RM instability, $a_0$ is the initial amplitude, $a_{ps}$ is the post-shock amplitude (taken as $a_0/4$, obtained from simulations), $\Delta u$ is the magnitude of the velocity jump, which is 16 $\mu m/ns$ according to simulations. Finally, decompression of the materials at the unstable interface increases the apparent mix width by nearly 30\% relative to the instability-only growth at $X_{Read}=600$ $\mu m$, illustrating the importance of this term in any planar laser-driven blast wave experiment. The dotted and dashed lines in Fig. \ref{fig:data} are the model results for $\left<\lambda\right>_0=15\mu m$ which represent the best-fit (dotted line) and the highest and lowest values of $\alpha_B$ (dashed lines) according to a $\chi^2$ test. In this case of initial perturbation, $G\sim3$ are being reached at $X_{Read}$$=$$540\mu m$ according to the model presented below, under the assumptions of $\alpha_B=0.03$ and 3D late dynamics. This should be the lower bound for $G$ at this time since the higher the value of $\alpha_B$, the higher the value of $G$ and also, $G$ is bigger in 2D than in 3D (2D bubbles are wider than in 3D \cite{oron2001dimensionality}). Due to the high bubble merger generation, $\left<\lambda\right>_0=15\mu m$ supplies the main results of this work.  In this case, the initial amplitudes in the model were set to $3-4\mu m$ for matching the experimental results. Note that this case was harder to machine and higher amplitudes than $1\mu m$ are possible.

We observe from Fig. \ref{fig:data} that our $\left<\lambda\right>=30\mu m$ data show a similar slope to the model result for $\alpha_{B}\sim0.03$. Focusing on the $\left<\lambda\right>=15\mu m$ case, which offers a better constraint, we get $\alpha_B=0.038\pm 0.008$. Note that applying the model with 2D or 3D parameters yield similar values of $\alpha_b$, so that either if the initial perturbation in the experiment is totally forgotten (3D) or the dynamics is still 2D but with $G>3$, this is the measured value of $\alpha_B$. This value is consistent with previously reported 2D simulations \cite{shimony2018density} and closer to 3D simulations than previous reported experiments.

\begin{figure}
\includegraphics[width = 3.375in]{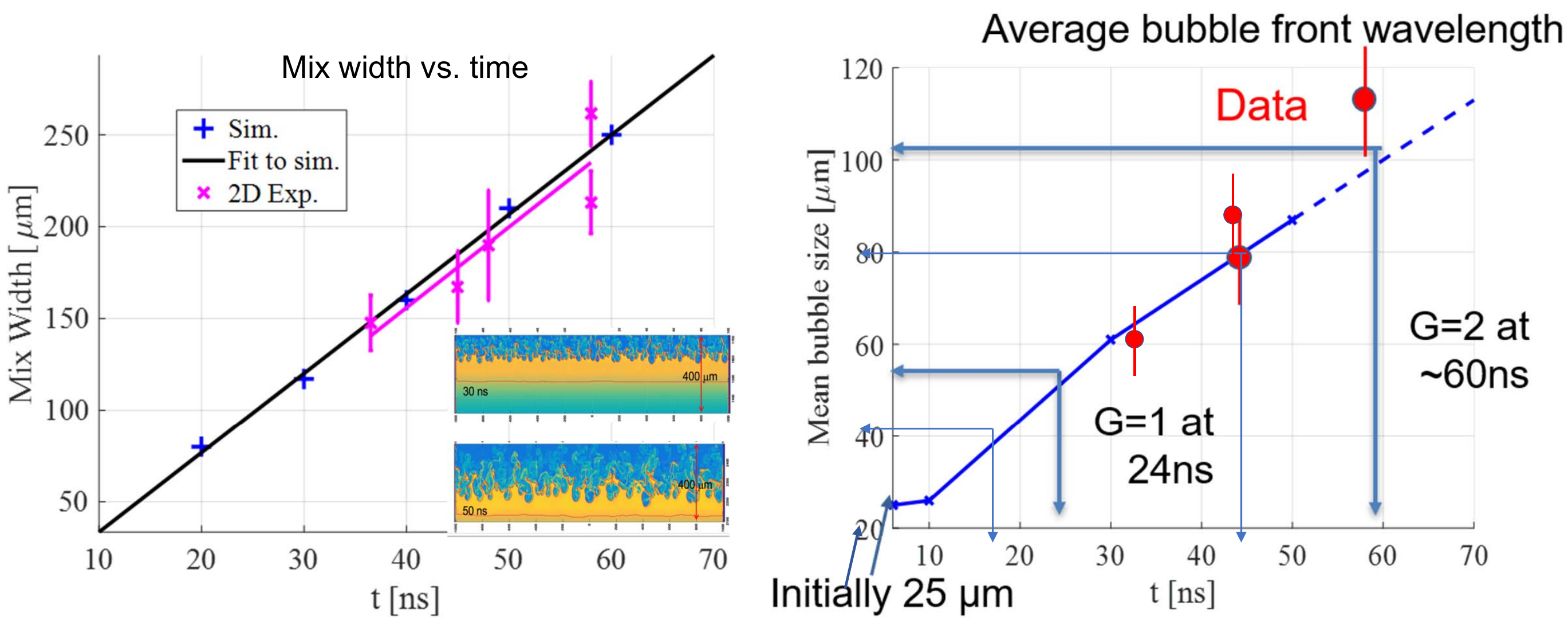}
\caption{(a) The mix width as a function of time in the 2D simulation and in the experimental results of a 2D initial perturbation. The insert presents pictures from the simulations at two times (30 ns and 50 ns) (b) The bubble size as a function of time for the same cases as (a).}
\label{fig:aBslope}
\end{figure}

In addition to the model, we performed a 2D Dafna \cite{malamud2014conceptual,wygoda2011relativistic} simulation which includes the evolution of the instabilities, while the trajectories of the shock and the interface, as well as the densities fit the LASNEX simulations. The simulation was preformed for the 2D $\left<\lambda\right>=30\mu m$ case, in which the initial perturbation was scanned and presented in the bottom of Fig. \ref{fig:target}. Fig. \ref{fig:aBslope} presents a good agreement between the Dafna simulation both for the width of the mixing zone (a) and for the average size of the bubbles (b). This agreement strengthens the model predictions, suggesting that RTI simulation and experiment can fit despite the discrepancies discussed above.

To summarize, we present a comprehensive NIF discovery science campaign of the self-similar RTI, evolving from controlled initial perturbations, which are similar to the ones used in previous simulations. The main result of this work is $\alpha_B=0.038\pm 0.008$. It was obtained by excluding the RMI and decompression effects on the growth of the mix width and retaining the RTI contribution. This result fits 2D simulations and closer to 3D simulations compared to previous experiments, resolving the known discrepancy between previous experiments and simulations. 

This work performed under the auspices of the U.S. Department of Energy by Lawrence Livermore National Laboratory under Contract DE-AC52-07NA27344, National Science Foundation through the Basic Plasma Science and Engineering program NSF 16-564, grant number 1707260,
and with support from UM-LLNL collaborative contract and from NRCN.

\bibliographystyle{apsrev4-2}   
\bibliography{DS_alpha_intro}

\end{document}